\documentclass[%
 aip,
 amsmath,amssymb,
 reprint,%
 floatfix,longbibliography]{revtex4-1}

\usepackage{graphicx}
\usepackage{dcolumn}
\usepackage{bm}
\usepackage{amsfonts}
\usepackage{mathtools}
\usepackage{braket}
\usepackage{multirow}
\usepackage{xcolor,soul}
\usepackage[utf8]{inputenc}
\usepackage[T1]{fontenc}
\usepackage{mathptmx}
\usepackage{etoolbox}
\usepackage{ulem, dsfont}

\makeatletter
\def\@email#1#2{%
 \endgroup
 \patchcmd{\titleblock@produce}
  {\frontmatter@RRAPformat}
  {\frontmatter@RRAPformat{\produce@RRAP{*#1\href{mailto:#2}{#2}}}\frontmatter@RRAPformat}
  {}{}
}%
\makeatother

\renewcommand{\inf}{\infty}

\newcommand{\mb}{\mathbf}

\definecolor{darkgreen}{rgb}{0.0, 0.5, 0.0}

\newcommand{\rev}[1]{{\color{black}#1}}

\begin{document}

\title[Decoupling band topology from criticality in bosonic systems]{Decoupling band topology from criticality in bosonic systems}

\author{Mariam Ughrelidze}
 \email{mariam.ughrelidze.gr@dartmouth.edu}
\affiliation{\mbox{Department of Physics and Astronomy, Dartmouth College, 6127 Wilder Laboratory, Hanover, NH 03755, USA}}

\author{Lorenza Viola}
\email{lorenza.viola@dartmouth.edu}
\affiliation{\mbox{Department of Physics and Astronomy, Dartmouth College, 6127 Wilder Laboratory, Hanover, NH 03755, USA}}

\author{Emilio Cobanera}
 \email{cobanee@sunypoly.edu}
\affiliation{\mbox{Department of Physics, SUNY Polytechnic Institute, 100 Seymour Avenue, Utica, NY 13502, USA}}
\affiliation{\mbox{Department of Physics and Astronomy, Dartmouth College, 6127 Wilder Laboratory, Hanover, NH 03755, USA}}

\date{\today}

\begin{abstract}
A new understanding of criticality in systems described by quadratic bosonic
Hamiltonians (QBHs) ties the emergence of long-range correlations to boundaries of dynamical, not thermodynamical, stability in parameter space. This separation occurs because the solution of the Heisenberg equations of motion is determined by an auxiliary {\it pseudo-Hermitian} dynamical system. The boundary points of a region of dynamical stability can be either exceptional points, generically associated with long-range correlations, or Krein collisions, where correlations can be either long- or short-range. We investigate the interplay of this landscape of possibilities with band topology and boundary physics, by relying on both specific examples and general arguments. The examples stem from a two-parameter, thermodynamically unstable family of QBHs obtained from the bosonic Su-Schrieffer-Heeger model by breaking particle conservation while preserving a chiral pseudo-symmetry. 
\rev{As a function of an interpolation parameter, distinct regions emerge within a fixed dynamical-stability phase, which prove to be topologically trivial and nontrivial, respectively.} The topological phase transition is a line of Krein collisions, which coincides with the closing of a band gap at zero and causes the localization length of the topologically mandated boundary zero modes to diverge before disappearing. We show that the chiral pseudo-symmetry induces, despite the broken particle-number symmetry, enough structure on its associated dynamical matrices to support a topological classification and a bulk-boundary correspondence, {\it independently} of dynamical stability. This strongly suggests that, for non-interacting bosons, topological physics extracted from basic index theory is insensitive to dynamical stability \rev{and, {\it a posteriori}, criticality.}

\end{abstract}

\maketitle

\section{Introduction}
\label{sec:intro}

The tenfold-way classification
of solid-state \rev{insulators and superconductors\cite{TopoInsul,TopoSuperc}} is one of the great accomplishments of condensed-matter physics. Narrowly construed, it is a topological classification of pseudo-symmetry~\cite{Ryu_2010} or many-body symmetry~\cite{Kitaev_2009, 2016Kennedy} classes of translationally-invariant, \rev{non-interacting fermionic systems, described by quadratic fermionic Hamiltonians (QFHs) in arbitrary
space dimension; that, however, is} not enough to account for its importance. In the following, we will use the term ``tenfold way" as a shorthand for a bigger piece of machinery including bulk-boundary correspondences~\cite{2016Chiu, Alldridge2020, 2023Alase} that link the tenfold-way classification to the electronic transport properties of \rev{bulk systems with terminations, whereby translational invariance is broken by boundaries. Interpreted in this broader sense,} 
the tenfold way explains why, at very low temperatures, a gapped system of independent fermions can behave as a conductor by way of gapless, boundary-localized fermions. It also predicts when exactly this can happen on very meager input: the spatial dimensionality of the system and some pseudo-symmetry\cite{Ryu_2010, 2016Chiu} or many-body symmetry\cite{Kitaev_2009, 2016Kennedy, Alldridge2020} considerations. Finally, the tenfold way predicts that the low-dimensional conductor that emerges at the boundary of a topological insulator or superconductor is special: unlike a conductor of the same spatial dimension achieved directly by confinement, a low-dimensional conductor on the boundary of a bulk-gapped topological material avoids Anderson localization (the metal-insulator transition driven by disorder)
and can support sharply quantized values of transport coefficients. The prototypical example of \rev{such a symmetry-protected topological phase}
is the integer quantum Hall effect, which provides the basis of the resistance standard in the revised international system of units today~\cite{PhysRevApplied.23.014025}. 

Another aspect of the tenfold way central to our investigation in this paper is its connection to quantum (zero temperature) phase transitions. The tenfold way predicts that, in a spatial dimension in which a symmetry class is non-trivial, the topological label of a gapped QFH in the class can only change at points where the many-body gap to the ground state closes. At the same time, for free fermions, the characteristic length of correlations in the ground state, which happens to always be a quasiparticle vacuum, diverges whenever the energy gap closes~\cite{Hastings}. Hence, within the framework of the tenfold way, topological \rev{quantum} phase transitions are automatically associated with fermionic criticality.

In this paper we explore some basic ideas, inspired by the work in Refs.~\onlinecite{2023Alase} and ~\onlinecite{Decon}, that should inform any attempt to uncover a ``bosonic $n$-fold way" of breadth and depth comparable to that of the \rev{fermionic} tenfold way, within the realm of closed, Hamiltonian systems~\cite{gardin2026manybodysymmetryprotectedzeroboundary}. Specifically, our main aim is to achieve \rev{a deeper}
theoretical understanding of the interplay between topological phase transitions in quadratic bosonic Hamiltonians (QBHs) and critical correlations in the bosonic quasiparticle vacuum (QPV)~\cite{Criticality}. Our findings strongly suggest that, \rev{in contrast with fermions,} bosons do not support
a tight link among changes in bulk topology, changes in boundary physics, and critical correlations in the QPV. This ``decoupling'' occurs because, unlike free fermions, the effective dynamics of free bosons is non-Hermitian~\cite{Decon} and this difference at the foundation propagates, often in unexpected ways, into more differences down the line. In our context, we find that the issue is not with the link between changes in bulk and boundary physics which, as we will see, mirrors free fermions well enough. The problem is \rev{rather} trying to link topological phase transitions to criticality: for bosons, such a connection is, if not clearly absent, at least puzzlingly hard to pin down.

Our evidence for this claim  comes from a thorough investigation of the class of bipartite QBHs, whereby two ``flavors'' of bosons are only coupled to each other. We find that we can characterize this class as a chiral pseudo-symmetry one regardless of space dimensionality, and that it supports a topological classification and {bulk-boundary correspondence~\cite{SchulzBaldesBBC}} in one dimension. We also find that band topology~\cite{BergholtzBulkBoundary, SpectralvsBand} and criticality in the QPV~\cite{Criticality} cleanly decouple for a prototypical family in this class. \rev{Specifically, this} family of translationally-invariant QBHs undergoes a topological transition (a change in the topological label), characterized by a diverging localization length \rev{of boundary modes} and a change in a bulk topological invariant, entirely \textit{within a single phase of dynamical stability} (more on this shortly) and \textit{without} long-range correlations being tied to the topological transition \rev{itself}. 
 
Most dynamical questions about a QBH for $n$ modes reduce to the spectral analysis of the $2n\times 2n$ pseudo-Hermitian dynamical matrix $\mb G$. Because $\mb G$ need not be Hermitian, QBHs split into two dynamical stability phases: stable ($\mb G$ is diagonalizable with a purely real spectrum) and {unstable\cite{RossignoliUnstable}} ($\mb G$ has complex eigenvalues or is non-diagonalizable), with the phase boundaries being marked by either {\textit{exceptional points} (EPs) or \textit{Krein collisions} (KCs) \cite{Decon,GohbergIndefiniteLA}.} These two kinds of spectral singularities have qualitatively different physical content. As established for general translationally invariant QBHs in Ref.~\onlinecite{Criticality}, EP-type degeneracies drive standard quantum-critical behavior in the QPV -- they engender long-range correlations, with a diverging correlation length. KC-type degeneracies, however, are more complex, and can be associated with critical, non-critical, or multicritical-like behavior.

The example we study in this manuscript is an interpolation model between two dimer QBHs. It is a close relative of the bosonic {Su-Schrieffer-Heeger (SSH) model~\cite{SSHOG,Ozawa2019}}, a one-dimensional (1D) model (a chain) with staggered nearest-neighbor hopping, and an additional pairing term that breaks particle-number conservation. As a function of two parameters $(s, \delta)$, the model exhibits two distinct phase boundaries: a topological transition line at $s = 1/2$ mediated by KCs, and a dynamical instability line at $\delta = 1$ mediated by EPs. The two lines are orthogonal in parameter space, and so are their diagnostics. The localization length of the boundary zero modes (ZMs) diverges at $s = 1/2$ and is entirely independent of $\delta$. QPV correlations diverge at $\delta = 1$, also independently of $s$. Furthermore, QPV covariance matrices are independent of $s$ throughout the dynamically stable regime. Finally, in this regime, the \textit{Krein gap}\cite{Criticality} vanishes identically, so the QPV is not unique. Therefore, any long-range correlations that may be supported are fixed by the choice of QPV, rather than by the topological transition. A bulk-boundary correspondence established via the index theorem for Toeplitz operators~\cite{lax2014functional} then captures the topological structure in both dynamically stable and unstable regimes.

The fact that a dynamically stable QBH can support nontrivial band topology is by itself well appreciated~\cite{Peano2015}. What is surprising about our model is that it supports multiple topological phases and transitions between them, all within a single phase of dynamical stability and, consistently with the general dichotomy of Ref.~\onlinecite{Criticality}, the resulting topological transition does \textit{not} carry the critical signatures in the same way one might expect based on free-fermion intuition. The emergence of long-range correlations is not guaranteed at the  topological transition and is conditional on the choice of QPV.

\section{A topological bosonic dimer model}
\label{sec:dyn-phases}

As we mentioned, we support our conclusions \rev{based on a combination of}
specific examples and general mathematical arguments. We will lead with the examples to keep the discussion as concrete as possible.

\subsection{From dimer QBHs to interpolation models}
\label{sec:dimer}

\begin{figure}[t]
    \centering
    \includegraphics[width=\linewidth]{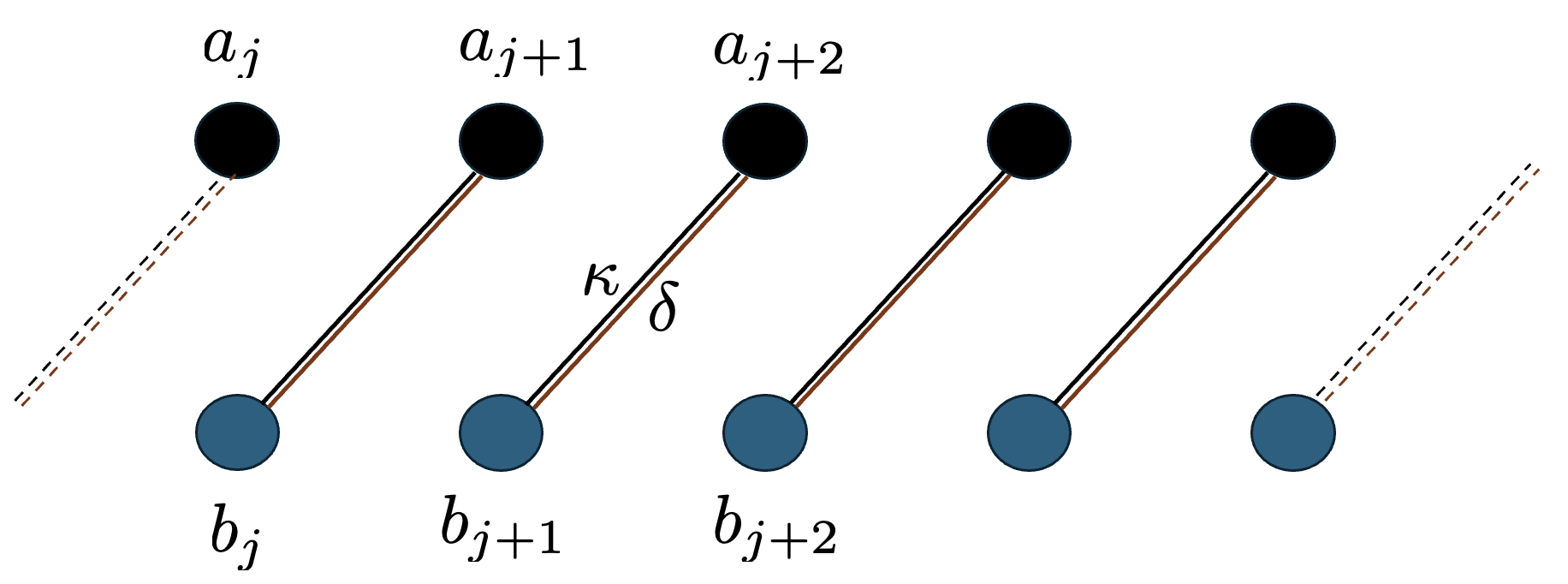}\vspace*{4mm}
    \includegraphics[width=\linewidth]{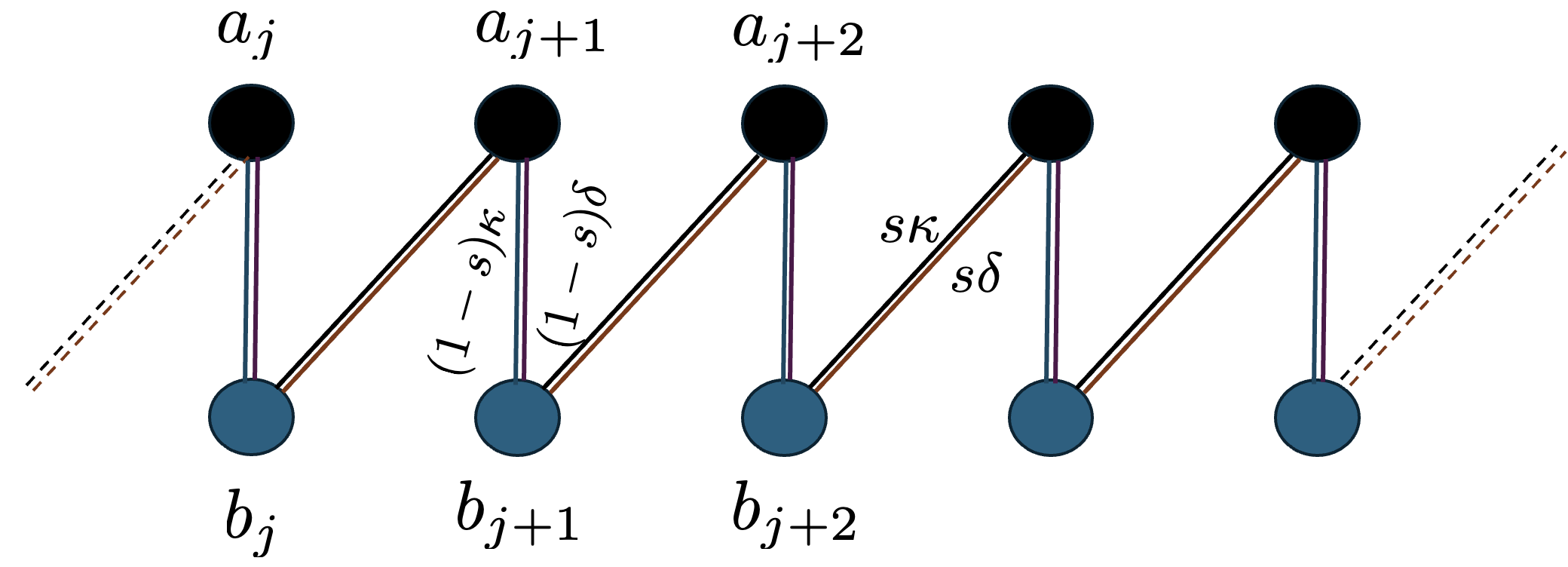}
    \caption{Pictorial representation of dimer and interpolation QBHs. Top: The dimer QBH ${\cal H}^{(2)}$ of Eq.\,\eqref{Eq: nthHamiltonian} ($n=1$). Bottom: The interpolation QBH ${\cal H}^{(2,0)}$ of Eq.\,\eqref{Eq: InterpolationHamiltonian}, seen as a continuous deformation of the
    $n=0$ (realized at $s=0$) into the $n=1$ (realized at $s=1$) dimer QBH. Dash lines are used to indicate that the chain is translationally invariant in either direction.}
    \label{fig:interpolationmodel}    
\end{figure}

Our example model can be described in terms of building blocks and interpolations between them. There are two flavors of bosonic creation and annihilation operators per lattice site, say, $a_j, a_j^\dagger$ and $b_j, b_j^\dagger$, with $j\in\mathbb{Z}$, see also  Fig.~\ref{fig:interpolationmodel} (top). The building-block QBHs are translationally-invariant dimer Hamiltonians labeled by an even integer:
\begin{align}
\label{Eq: nthHamiltonian}
    {\cal H}^{(2n)}\equiv \sum_{j}
    \left(\kappa\, a_{j+n}^\dagger b_j
        + \delta\, a_{j+n}^\dagger b_j^\dagger + \text{H.c.}\right), \quad 
\kappa,\delta\geq 0.        
\end{align}
The building blocks thus feature two continuous parameters $\kappa,\delta$ and one integer parameter $n$. As Fig.~\ref{fig:interpolationmodel} makes it clear, $n$ stands for a relative displacement: all the building-block Hamiltonians can be obtained from ${\cal H}^{(0)}$ by displacing the $b$ degrees of freedom $n$ sites to the left with respect to the $a$ degrees of freedom. This explains some exact spectral features of our interpolation models below. \rev{Following Ref.\,\onlinecite{DynamicalMetastability}, we define the following three types of boundary conditions (BCs):}
\begin{itemize}
    \item \textit{Open BC} (OBCs): A finite chain of lattice sites with a hard-wall boundary on each \vspace*{-1.5mm}side. 
    \item \textit{Bi-infinite BC} (BIBCs): A chain extending infinitely in both directions, with no \vspace*{-1.5mm}boundaries.
    \item \textit{Semi-infinite BC} (SIBCs): A pair of disjoint chains, each extending infinitely in one direction, to the left or the right, with a boundary on the other end.
\end{itemize}

The main characteristics of the building-block QBHs are flat bands for BIBCs and $2n$ exact, perfectly localized bosonic ZMs per boundary for OBCs. Our goal is to understand whether the ZMs are topologically mandated and symmetry-protected. However, these questions require careful consideration as the bands can be real or complex-valued. This motivates us to explore the following interpolation model
\begin{align}
\label{Eq: InterpolationHamiltonian}
    {\cal H}^{(2,0)}(s) \equiv s\, {\cal H}^{(2)} + (1-s)\, {\cal H}^{(0)}, \quad
    0\leq s\leq 1,
\end{align}
as a function of $\delta$ and $s$ to test the relationship between bulk and boundary
physics. This is the interpolation QBH alluded to in the title of this section. For $\delta=0$, our interpolation model is a special case of the celebrated bosonic SSH chain~\cite{Ozawa2019}: The bosonic SSH model is characterized by staggered hopping and particle-number conservation. \rev{Importantly, in our 
model, particle-number conservation is broken by an additional staggered bosonic pairing term. In the following, we will fix $\kappa = 1$ and focus on $\delta > 0$, unless otherwise specified. From an implementation standpoint, it is worth noting that all the coupling terms entering our model QBH are becoming increasingly available across different experimental platforms; in particular, recent realizations of related bosonic-chain models have been reported in a nano-optomechanical network~\cite{2024Slim} and a superconducting parametric cavity \cite{2024Busnaina}, with additional possibilities emerging in parametric trapped-ion arrays~\cite{trappedions} and magnonic systems~\cite{Wang,gardin2026manybodysymmetryprotectedzeroboundary}.}

Building on standard methods~\cite{Blaizot,Decon}, let us collect the creation and annihilation operators in a bosonic Nambu array
\begin{equation}
\label{eq: Nambuordering}
\Phi\equiv [a_1,\ldots,a_N,b_1,\ldots,b_N,a_1^{\dag},\ldots,a_N^{\dag}, b_1^{\dag},\ldots,b_N^{\dag}]^T,
\end{equation}
so that, given a QBH $\mathcal{H}$, we may write 
\begin{equation}
    \mathcal{H}=\frac{1}{2}\Phi^{\dag}\mb H\Phi ,
\end{equation}
for some $4N\times 4N$ Hermitian matrix $\mb H$. After setting $\hbar=1$, the Heisenberg equations of motion may be cast in the form 
\begin{align}
\label{Eq: EOMs}
i\,\frac{d\Phi}{dt}=\boldsymbol{\tau}_3\mb H \Phi\equiv \mb G\Phi,
\quad \boldsymbol{\tau}_i\equiv \boldsymbol{\sigma}_i\otimes\mb 1 ,
\end{align}
where $\boldsymbol{\sigma}_i$ are auxiliary
Pauli matrices, and $\mb G$ is the bosonic \textit{dynamical matrix} \cite{Decon,DynamicalMetastability}. For the building-block Hamiltonians of Eq.~\eqref{Eq: nthHamiltonian}, we may write 
\begin{align}
\mb G^{(2n)}\equiv \begin{pmatrix}
    \mb K^{(2n)} & \boldsymbol{\Delta}^{(2n)}\\
    -\boldsymbol{\Delta}^{*(2n)} & -\mb K^{*(2n)}
\end{pmatrix},
\end{align}
in terms of
\begin{align*}
\mb K^{(2n)} =
    \begin{pmatrix}
        0 & \mb F^{\dag n}\\
        \mb F^{n} & 0
    \end{pmatrix}, \quad
\boldsymbol{\Delta}^{(2n)}= \delta
    \begin{pmatrix}
        0 & \mb F^{\dag n}\\
        \mb F^{n} & 0
    \end{pmatrix}.
\end{align*}
The operator $\mb F$, specified as follows, 
\begin{align}
\label{Sshift}
    \mb F&\equiv \begin{cases}
                \begin{array}{llll}
                \mb T_N &\!\!\equiv & \sum\limits_{j=1}^{N-1}\vec{e}_j\vec{e}_{j+1}^{\,\dag} & \text{OBC},\\ 
                \mb V&\!\!\equiv &\sum\limits_{j=-\infty}^{\infty}\vec{e}_j\vec{e}_{j+1}^{\,\dag} & \text{BIBC},\\ 
               { \mb T}&\!\!\equiv & \mb T^L+\mb T^R
               &\text{SIBC}.\\
                \end{array}
              \end{cases},
\end{align} 
imposes the BCs. Here,  $\vec{e}_j$ is the $j$-th canonical basis vector in ${\mathbb C}^N$, {and the Toeplitz operators 
\begin{equation}
    \mb T^L\equiv \sum\limits_{j=1}^{\infty}\vec{e}_j\vec{e}_{j+1}^{\dag}, 
    \quad \mb T^R\equiv \sum\limits_{j=-\infty}^{-1}\vec{e}_j\vec{e}_{j+1}^{\dag}
    \label{LR}
\end{equation}
describe the half-infinite chain retaining a single left- (L-SIBC) or right- (R-SIBC) boundary, respectively. While defining SIBCs as in Eq.~\eqref{Sshift} is essential for rigorously connecting the spectrum of an infinite-size system to the one of a finite system under OBCs, a simpler (and more standard) notion solely based on L- (R-)SIBC will suffice in the present context of topological classification, as we shall see (Sec.~\ref{sec: beyondstable}).}

\subsection{Bulk dynamical regimes}
\label{sec:dynamical-phases}

The dynamical matrix of a QBH is not Hermitian unless the bosonic pairing terms vanish. It is, nonetheless, {\textit{pseudo-Hermitian}\cite{PseudovsPT,KawabataSato}} with respect to the indefinite metric $\bm{\tau}_3$, i.e., $\mathbf{G}^\dag = \bm{\tau}_3 \mathbf{G}\bm{\tau}_3$. It also satisfies a charge-conjugation constraint $\boldsymbol{\tau}_1\mb G\boldsymbol{\tau}_1=-\mb G^{*}$ \cite{Decon}. Since $\mb G$ is generically non-Hermitian, there is no guarantee that observables will undergo bounded motion. Therefore, one can define a notion of \textit{dynamical stability}, whereby the expectation value of any observable remains bounded for all time, independently of the choice of initial state. For a QBH, a necessary and sufficient condition for dynamical stability is that the associated dynamical matrix is diagonalizable, with a purely real spectrum \cite{Decon}. Under this condition, the time evolution of the creation and annihilation operators is strictly bounded [as seen from Eq.~\eqref{Eq: EOMs}], which, for unitary dynamics, in turn guarantees that the expectation values of arbitrary observables remain bounded.

\begin{figure*}[t]
    \centering
    \includegraphics[width=0.9\linewidth]{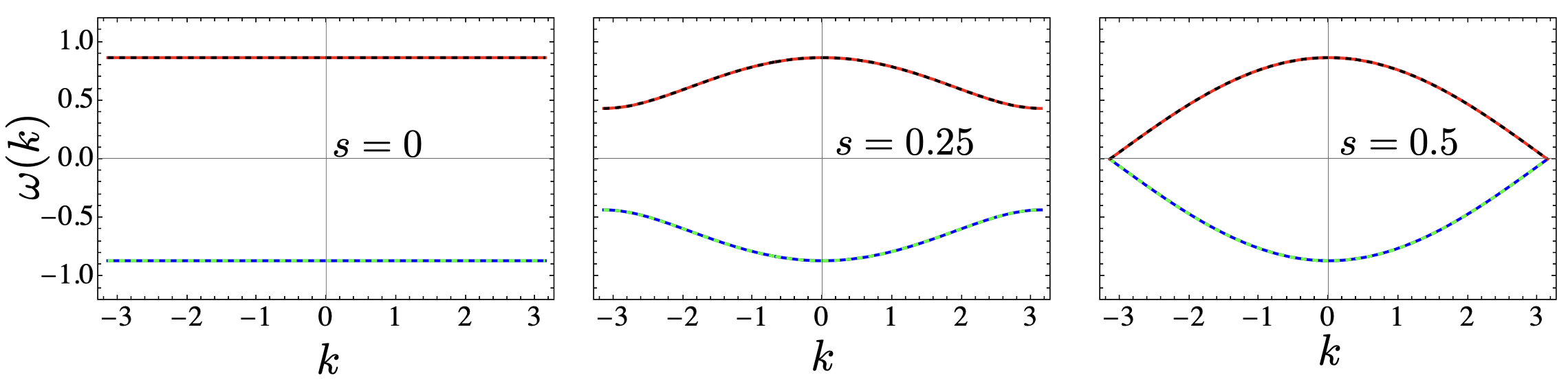}
    \caption{{Band structure of the interpolation model ${\cal H}^{(2,0)}$, $\omega_{j,\pm}(\delta,s,k)$ of Eq.\,\eqref{eq:omegajs} for $\delta=0.5$. 
    Only $s$ values up to $0.5$ are depicted, as the bands are symmetric under $s\mapsto1-s$.}}
    \label{fig:bands}
\end{figure*}

In the bulk (under BIBCs),  translational invariance allows to extract the \textit{Bloch dynamical matrix} in momentum space, $\mb g(\bm k)$, which, in turn, satisfies $\mathbf{g}^\dag(\bm{k}) = \bm{\tau}_3 \mathbf{g}(\bm{k})\bm{\tau}_3$ and $\mathbf{g}^*(\bm{k}) = -\bm{\tau}_1 \mathbf{g}(-\bm{k}) \bm{\tau}_1$. In the dynamically stable phase, the eigenstructure of $\mb g(k)$ takes the following form\cite{Criticality}:
\begin{align}
    \mathbf{g}(\bm{k})\vec{v}_{j,\pm}(\bm{k}) &= \pm \omega_j(\pm \bm{k}) \vec{v}_{j,\pm}(\bm{k})
\end{align}
with $\vec{v}_{j,\varsigma}^\dag(\bm{k})\bm{\tau}_3 \vec{v}_{j',\varsigma'}(\bm{k}) =\varsigma \delta_{jj'}\delta_{\varsigma \varsigma'}$ and $ \vec{v}_{j,\mp}(\bm{k}) = \bm{\tau}_1 \vec{v}_{j,\pm}^*(-\bm{k})$. Here $\varsigma=\pm 1$ denotes the {\textit{Krein signature} of these eigenvectors \cite{GohbergIndefiniteLA}}. Given a vector 
$\vec{v}$, its Krein signature is defined as the sign of $\vec{v}^\dag \bm{\tau}_3\vec{v}$, with the convention $\text{sgn}(0)=0$. The eigenvectors of a dynamically stable dynamical matrix satisfy a normalization condition:
\begin{align}
  \varsigma\equiv   \vec{v}_{j,\pm}^{\dag}\boldsymbol{\tau}_3\vec{v}_{j,\pm} =\pm 1.
\end{align} When different from zero, the Krein signature
distinguishes vectors associated with bosonic quasiparticle creation operators (one signature) from vectors
associated with annihilation operators (the opposite signature). Quadratures (position and momentum-like operators), by contrast, are associated with null vectors.

The dynamical stability phase boundaries of QBHs are comprised of two types of spectral degeneracies, EPs and KCs. If the dynamical matrix possesses an EP or a KC, its spectrum can be made complex by an arbitrarily small generic perturbation. {At an EP, two or more eigenvalues and corresponding eigenvectors coalesce and the dynamical matrix becomes non-diagonalizable\cite{HeissEP,BergholtzEP,WiersigEP}}. At a KC, the dynamical matrix has an eigenvalue with at least two corresponding eigenvectors of opposite Krein signatures. 

The Fourier transform of $\mb G^{(2n)}$ under BIBCs yields: 
\begin{align*}
    \mb g^{(2n)}(k)&=\begin{pmatrix}
    \mb K^{(2n)}(k) & \boldsymbol{\Delta}^{(2n)}(k)\\
    -\boldsymbol{\Delta}^{*(2n)}(-k) & -\mb K^{*(2n)}(-k)
\end{pmatrix}, \quad \text{with} 
\end{align*}
\begin{align*}
\mb K^{(2n)}(k)=\begin{pmatrix}
   0 & e^{-ink}\\
   e^{ink} & 0
\end{pmatrix}, \quad 
\boldsymbol{\Delta}^{(2n)}(k) =\delta\begin{pmatrix}
    0 & e^{-ink}\\
    e^{ink} & 0
\end{pmatrix}.
\end{align*}
The Bloch dynamical matrix of the interpolation Hamiltonian in
Eq.~\eqref{Eq: InterpolationHamiltonian} is
\begin{align*}
    \mb g^{(2,0)}(k)=(1-s)\mb g^{(0)}(k)+s\,\mb g^{(2)}(k),
\end{align*}
that is, an interpolation of the Bloch dynamical matrices.

Before proceeding to the band structure, we record a symmetry and a pseudo-symmetry of the interpolation model. First, the dynamical matrix commutes with the matrix
\begin{equation}
\label{eq:S-def}
\mb S \;\equiv\;
\begin{pmatrix} 0 & i\boldsymbol{\sigma}_z \\
                i\boldsymbol{\sigma}_z & 0 \end{pmatrix}, \quad {\mb S}{\mb S}^\dagger = \mb 1. 
\end{equation}
{Since $\mb S$ has the structure of a bosonic dynamical matrix, $\mb S^{\dag}=\boldsymbol{\tau}_3\mb S\boldsymbol{\tau}_3$ and $\mb S^{*}=-\boldsymbol{\tau}_1\mb S\boldsymbol{\tau}_1$,
it can be regarded as descending from a Hermitian quadratic form that commutes with the interpolation model, that is, a many-body symmetry. The parent Hamiltonian of $\mb S$ can be shown to be $\mathcal H_S=\frac{i}{2}\sum_j\left((a_j^{\dag 2}-a_j^2)-(b_j^{\dag 2}-b_j^2)\right),$ which generates a squeezing transformation.} 
In addition, the dynamical matrix anti-commutes with the unitary
matrix
\begin{equation}
\label{eq:R-def}
\mb R \;\equiv\; \mathrm{diag}(-1,+1,-1,+1), \quad \mb R^2 = \mb 1.
\end{equation}
Also, $[\mb R, \mb S] = 0$ and $\{\mb S, \boldsymbol{\tau}_3\} = 0$, while $[\mb R,\boldsymbol{\tau}_3] = 0$. The anti-commutation relation $\{\mb R, \mb g(k)\} = 0$ {cannot be lifted to a many-body symmetry\cite{gardin2026manybodysymmetryprotectedzeroboundary}, but instead identifies $\mb R$ as a chiral 
pseudo-symmetry of the interpolation model}. It reduces to the usual chiral symmetry
of the bosonic SSH model for $\delta=0$. Since $\mb R$ commutes with $\boldsymbol{\tau}_3$, it preserves the Krein signature. Combined with $\{\mb R, \mb g(k)\} = 0$, this means $\mb R$ pairs eigenvectors with positive Krein signature at eigenvalue $\omega$ with positive Krein signature eigenvectors at eigenvalue $-\omega$, and similarly for the negative Krein signature eigenvectors.

The role of $\mb S$ is complementary. The commutation
$[\mb S, \mb g(k)] = 0$ implies that $\mb S$ preserves the
eigenspaces of $\mb g(k)$, while $\{\mb S, \boldsymbol{\tau}_3\} = 0$
implies that $\mb S$ flips the Krein signature.
Together, if $\vec{v}$ is an eigenvector of $\mb g(k)$ with a positive Krein signature
at eigenvalue $\omega$, then $\mb S\vec{v}$ is a negative Krein signature
eigenvector at the same eigenvalue. So every band of $\mb g(k)$ is
doubly degenerate, with the two eigenvectors carrying opposite
Krein signatures.

We now compute them explicitly. We have a pair of twofold degenerate bands (see Fig.~\ref{fig:bands}): 
\begin{align}
\label{eq:omegajs}
\omega_{1,+}(\delta,s,k) &= \omega_{2,-}(\delta,s,k) \\
\nonumber & 
 =-\sqrt{(\delta^2-1)\bigl[2s(s-1)(\cos k - 1) - 1\bigr]}, \\
 \omega_{2,+}(\delta,s,k) &= \omega_{1,-}(\delta,s,k) \\
\nonumber &= 
+\sqrt{(\delta^2-1)\big[2s(s-1)(\cos k -1) - 1\bigr]}.
\end{align}
Here the index $j=1,2$ labels the different bands, whereas $\pm$ denotes Krein signature. The band gap then reads 
\begin{align}
  \Delta_{\text{band}} & \equiv\underset{k}{\text{min}} |\omega_{1,+}(\delta,s,k)-\omega_{2,+}(\delta,s,k)| \notag\\ 
  &=2|1-2s|\sqrt{1-\delta^2}.
  \label{bgap}
\end{align}
Unsurprisingly, the lines at which the band gap closes are physically relevant as we will see shortly. The regimes of dynamical stability and instability of ${\cal H}^{(2,0)}$ can be diagnosed directly from the band structure. For $0 \leq \delta < 1$, $\mb g(k)$ is diagonalizable, the bands $\omega_{j,\pm}(\delta, s, k)$ are all purely real, and the model is dynamically stable for all $s$. At $\delta = 1$, the model develops an EP and $\mb g(k)$ ceases to be diagonalizable. For $\delta > 1$, the bands become complex and, accordingly, the system is dynamically unstable. 

\begin{figure}[t]
    \centering
    \includegraphics[width=\linewidth]{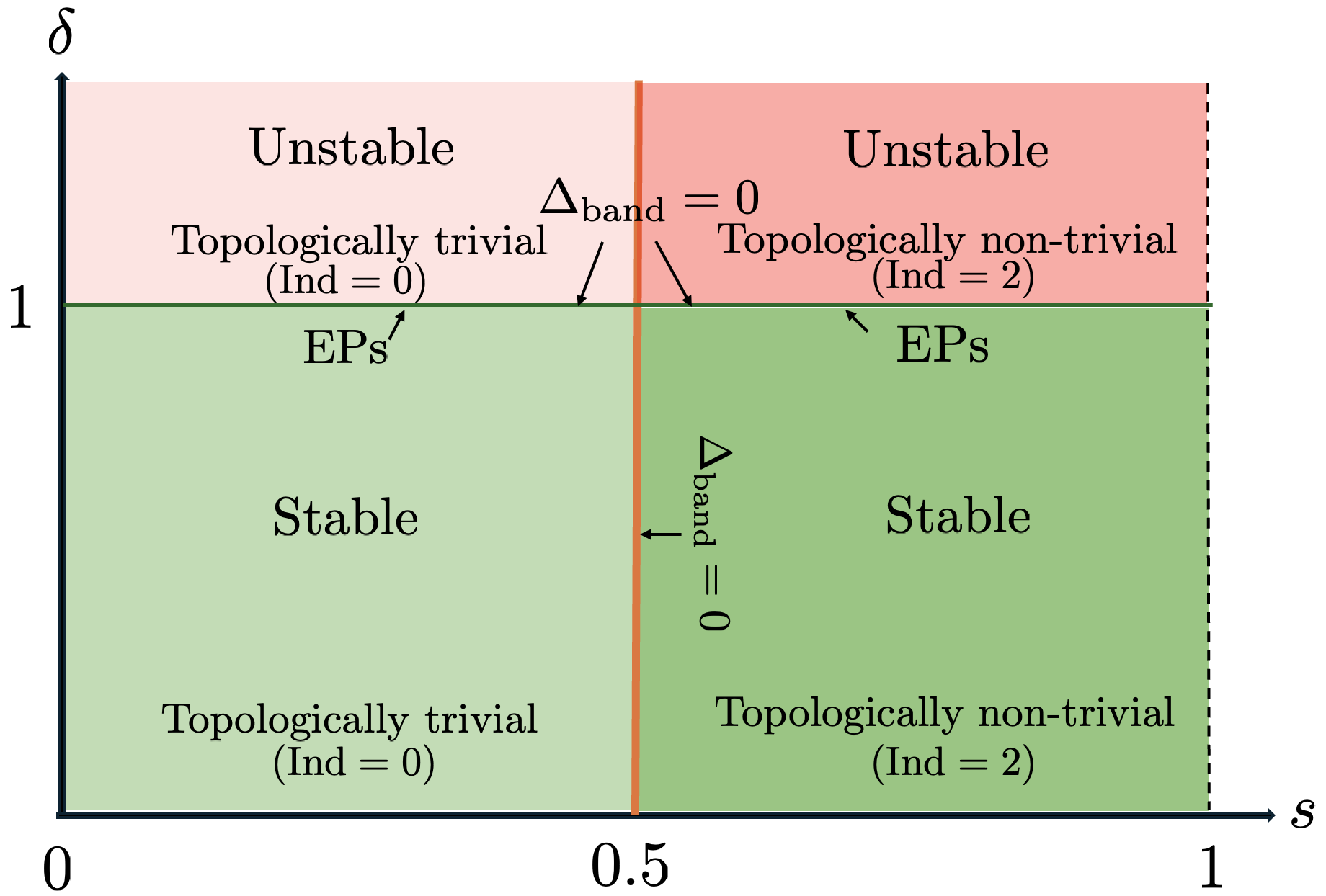}
    \caption{Dynamical stability and topological phase diagram of the interpolation model ${\cal H}^{(2,0)}$, showing both a topological phase transition and a dynamical-stability transition. For any fixed $\delta$, a topological phase transition occurs at $s=1/2$, where the band gap closes,
    while a dynamical stability transition happens as the line of EPs at $\delta=1$ is crossed. In the dynamically stable phase, the Krein gap is zero throughout, due to KCs. The dynamically stable (unstable) phases are also marked in green (salmon) color. Light shades correspond to topologically trivial regions \rev{with zero index}, while the dark shades correspond to topologically non-trivial \rev{ones with index 2 (as defined in Eq.\,\eqref{index})}. 
    }
    \label{fig:StabilityDiagram}
\end{figure}

In the dynamically stable phase, the $\boldsymbol{\tau}_3$-normalized and positive Krein-signature eigenvectors corresponding to the bands $\omega_{j,+}$ are given by
\begin{align}
\label{Eq: eigenvecs}
\vec{v}_{1,+}(\delta,s,k) \nonumber
& = \sqrt{\frac{\delta^2/2}{1-\delta^2}}
   \left(
   \frac{ \omega_{1,-}}
        {\delta [1+ s(-1 + e^{ik})]},
   \;-\frac{1}{\delta},\;0,\;1
   \right)^{T}, \\
\vec{v}_{2,+}(\delta,s,k)
& = \sqrt{\frac{\delta^2/2}{1-\delta^2}}
   \left(
   \frac{ \omega_{2,-}}
        {\delta [1+ s(-1 + e^{ik})]},
   \;-\frac{1}{\delta},\;0,\;1
   \right)^{T}.
\end{align}
The negative-Krein-signature partners are obtained as 
$
\vec{v}_{j,-}(\delta,s,k) = \boldsymbol{\tau}_1\,\vec{v}_{j,+}(\delta,
s, -k)^{*},
$
thanks to the constraint $\mb g(k)^{*} = -\boldsymbol{\tau}_1\,\mb g(-k)\,\boldsymbol{\tau}_1$ satisfied by
any bosonic dynamical matrix. One can
verify directly from Eq.~\eqref{Eq: eigenvecs} that the symmetry $\mb S$
acts on these eigenvectors as $\mb S\,\vec{v}_{1,+} = -i\vec{v}_{2,-}$ and $\mb S\,\vec{v}_{2,+} = -i\vec{v}_{1,-}$.

Since every eigenspace of $\mb g(k)$ is two-dimensional and spanned by eigenvectors of opposite 
Krein, that is, $\boldsymbol{\tau}_3$-signature $(+,-)$, for each $\omega_{j,+}$ we can construct 
a family of eigenvectors with a positive Krein signature parametrized by $(\eta(k), \theta(k)) \in 
\mathbb{R} \times S^1$. For $\omega_{1,+}$ the family is
\begin{align}
\label{eq:family}
\vec{u}_{1,+}(\delta,s,k;\eta,\theta)
\nonumber& =
  \cosh\left(\eta(k)\right)\;\vec{v}_{1,+}(\delta,s,k) 
  \\&
  +
  e^{i\theta(k)}\sinh\left(\eta(k)\right)\;\vec{v}_{2,-}(\delta,s,k),
\end{align}
{where $(\eta(k),\theta(k))$ set a
momentum-resolved two-mode squeezing amplitude and phase. To see that $\vec{u}_{1,+}$ has $+1$ Krein signature, note that $\vec{v}_{1,+}^{\dag}\boldsymbol{\tau}_3\vec{v}_{1,+}=1$, $\vec{v}_{2,-}^{\dag}\boldsymbol{\tau}_3\vec{v}_{2,-}=-1$, and $\vec{v}_{1,+}^{\dag}\boldsymbol{\tau}_3\vec{v}_{2,-}=0$. Therefore, 
\begin{align*}
    \vec{u}_{1,+}^{\dag}\boldsymbol{\tau}_3\vec{u}_{1,+}&=(\cosh\eta)^2\vec{v}_{1,+}^{\dag}\boldsymbol{\tau}_3\vec{v}_{1,+}+(\sinh\eta)^2\vec{v}_{2,-}^{\dag}\boldsymbol{\tau}_3\vec{v}_{2,-}\\&=(\cosh\eta)^2-(\sinh\eta)^2=1 ,
\end{align*}
yielding a definite Krein signature. The corresponding eigenvector} at $\omega_{2,+}$ is, 
\begin{align}
\label{eq:family2}
    \vec{u}_{2,+}(\delta,s,k;\eta,\theta) = \mb R\,\vec{u}_{1,+}(\delta,s,k;\eta,\theta),
\end{align}
determined by the chiral pseudo-symmetry.
The negative-Krein-signature partners are defined by the relation
$\vec{u}_{j,-}(k;\eta,\theta) = \boldsymbol{\tau}_1\,\vec{u}_{j,+}(-k;\eta,\theta)^*$. {A consequence that will later become important pertains to the \textit{Krein gap} introduced in Ref.~\onlinecite{Criticality} as a quantity responsible for controlling the decay of correlations in dynamically stable QBHs that may lack a ground state. The Krein gap is a generalization of the many-body spectral gap and quantifies the separation between bands of opposite Krein signature. Crucially, it signals how close a given system is to a dynamical stability boundary. By using the definition 
\begin{equation}
\Delta_\text{Krein} \equiv \min_{k,j,j'} \big|\omega_j(k) + \omega_{j'}(-k)\big|,
\label{KG}
\end{equation}
one can see that the Krein gap is \textit{identically zero} throughout
the dynamically stable region of this model, due to the degeneracy of $\pm 1$ Krein signature bands, and thus sits at the cusp of dynamical stability.} 

\rev{The resulting dynamical-stability phase diagram, along with the topological characterization from Sec.~\ref{sec: beyondstable}, is summarized in Fig.~\ref{fig:StabilityDiagram}.
}

\subsection{Boundary physics}
\label{sec:boundary}

\begin{figure*}[t]
    \centering
    \includegraphics[width=0.8\linewidth]{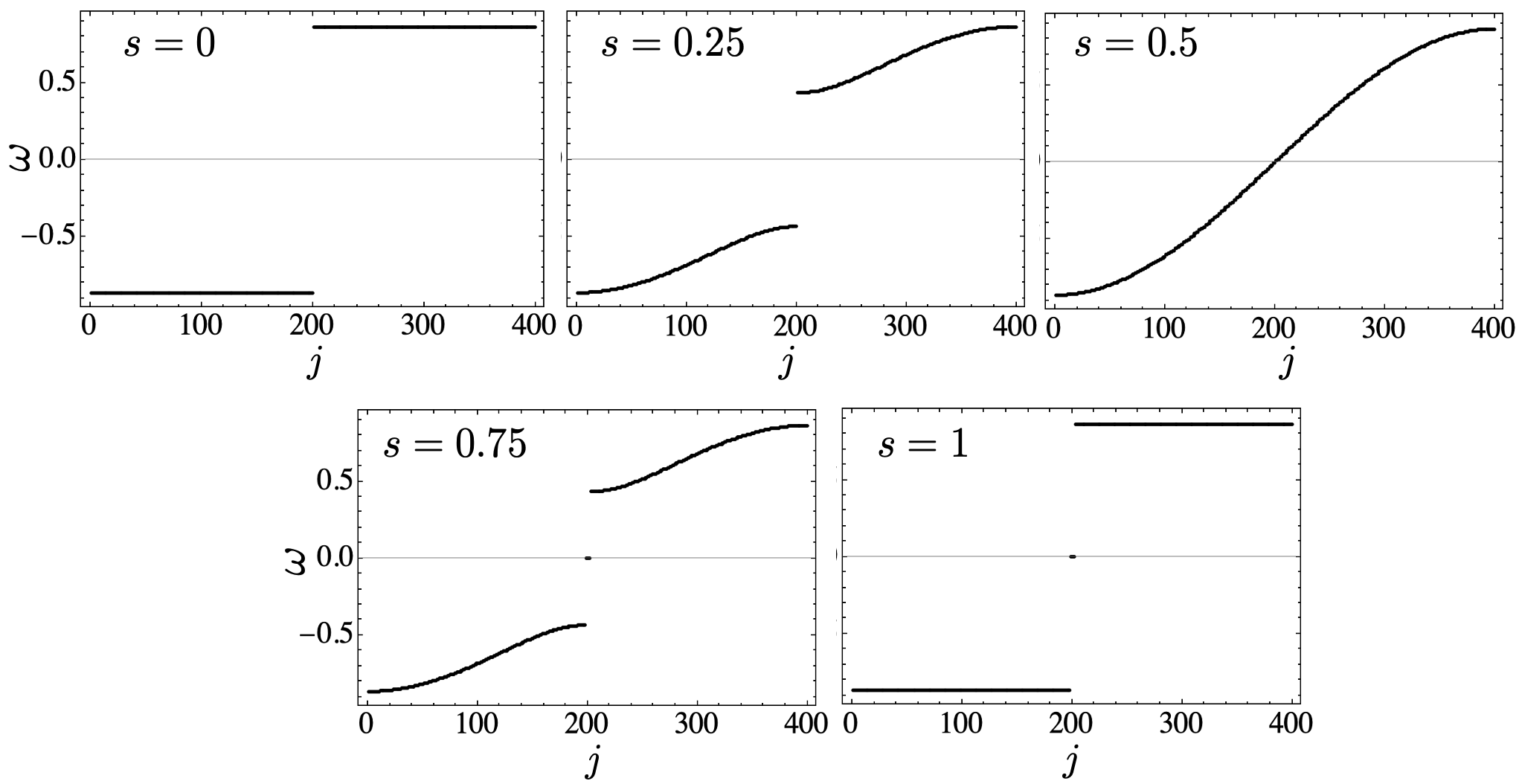}
    \caption{The spectrum of the dynamical matrix of the model Hamiltonian ${\cal H}^{(2,0)}$ in Eq.\,\eqref{Eq: InterpolationHamiltonian} under OBCs for $\delta=0.5$ and $N=100$. 
    Zero-energy boundary modes appear for $s\geq 1/2$.}
    \label{fig:zeromodes}
\end{figure*}

\begin{figure*}[t]
    \centering
    \includegraphics[width=0.8\linewidth]{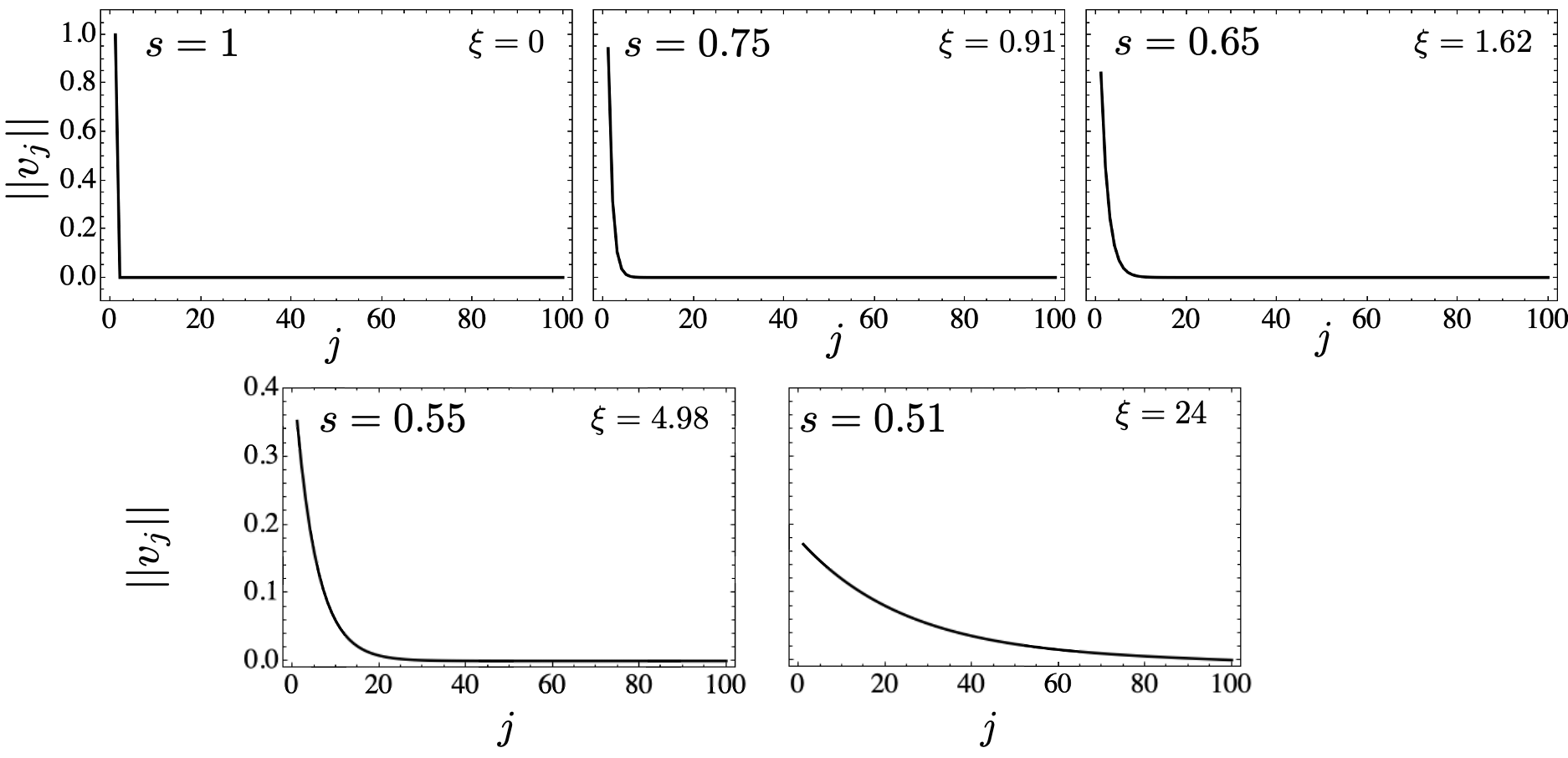}
    \vspace*{-3mm}
    \caption{The localization length $\xi$ of the {left} boundary ZMs diverges as the interpolation parameter approaches the topological phase
    transition. {Right boundary ZMs have the same localization length and undergo an identical behavior.} Parameters are chosen as in Fig.\,\ref{fig:zeromodes}.}
    \label{fig:Localization}
\end{figure*}

\rev{The interpolation model is, at $\delta = 0$, the bosonic SSH model (up to reparameterization), which undergoes a topological phase transition at $s = 1/2$ characterized by an integer-valued winding number}. A natural question is what becomes of this topology once a particle-number non-conserving pairing term is introduced.
As $s$ increases from 0 to 1, the interpolation model deforms from a chain where the dominant couplings are inside each unit cell to one where they are across unit cells. In the bosonic SSH model this transition leaves unpaired degrees of freedom at the boundaries of the open chain, producing edge-localized zero energy modes. We now look for the analogous effect in our model by examining the OBC spectrum.
For bosonic QBHs, changing the BCs can, in principle, alter the dynamical stability of the system altogether~\cite{DynamicalMetastability}. However, for the interpolation model studied here, we numerically confirm that the dynamical stability phase diagram is identical under bulk and OBCs. Additionally, the OBC dynamical matrix displays a spectral gap around zero for $s<1/2$. 
At $s=1/2$ this gap closes. For $s>1/2$, it reopens, but in contrast with the $s<1/2$ phase, modes at zero energy emerge in the gap; see Fig.\,\ref{fig:zeromodes}. Inspection of the eigenvectors corresponding to these ZMs reveals that they are exponentially localized at the boundaries of the chain, that is, $\psi_j \propto \exp(-j/\xi)$ near the left edge and $\psi_j \propto \exp(-(N-j+1)/\xi)$ near the right edge, where $\xi$ is the localization length. This characteristic length $\xi$ diverges as $s\rightarrow 1/2^{+}$; see Fig.\,\ref{fig:Localization}. The divergence of the localization length is a hallmark of topological transitions, analogous to the behavior of the edge states in the SSH model at its critical point. 
\rev{The topological nature of the transition will be put on more solid footing in Sec.~\ref{sec: beyondstable}}.

\rev{Having numerically computed the boundary ZMs under OBCs, we now obtain them analytically in closed form in the limit $N\rightarrow \infty$, i.e., under semi-infinite BCs.} For concreteness, we will {first compute the left ZMs, by considering the half-infinite chain that retains a left boundary.} To build the ZMs, we seek a linear combination of bosonic creation operators, $A_L =\mathcal{N}^{-1}\sum_{j} c_j\, a_j^{\dag}$, that commutes with the Hamiltonian, that is, $[\mathcal{H}(s), A_L]=0$. {This ansatz comes from the fact that the open boundary on the left leaves an unpaired $a$-mode.} Let us calculate 
\begin{widetext}
\begin{align}
    [\mathcal{H}^{(2,0)}(s),a_j^{\dag}]=
    \begin{cases}
      (1-s)b_j^{\dag}+(1-s)\delta b_j  & j=1\\
      (1-s)b_j^{\dag}+(1-s)\delta b_j+s b_{j-1}^{\dag}+s\delta b_{j-1}
      & j>1.
    \end{cases}
\end{align}
\end{widetext}
Now we can make the crucial observation that, in the absence of a right boundary, the $b$-operator contributions from neighboring sites can be made to cancel telescopically and
 \begin{align}
    \label{al_mode}
    A_L=\frac{1}{{\cal N}}\sum_{j=1}^{\infty} a_j^{\dag}
        \left(-\frac{1-s}{s}\right)^{j-1}
\end{align}   
commutes with ${\cal H}^{(2,0)}$. One also learns that $A_L^\dagger$
commutes with the Hamiltonian as well. Letting
\begin{align*}
    {\cal N}^2 \equiv \frac{s^2}{2s-1},
\end{align*}
the commutator of these two ZMs reads
\begin{align*}
    [A_L^{\dag},A_L]&=\frac{1}{{\cal N}^2}\sum_{j=1}^\infty
        \left(-\frac{1-s}{s}\right)^{2(j-1)}\nonumber
    =\begin{cases}
    1 & s>1/2\\
    \inf & s<1/2.
\end{cases}
\end{align*}
In particular, it is finite only for the $s>1/2$ region of the dynamically stable phase, where the boundary ZMs constitute a conjugate pair of canonical bosons.

{An entirely similar construction can be carried out for the half-infinite chain that retains a right boundary. The boundary ZMs of this half-chain are now built out of $b$-type creation operators. Explicitly, the relevant commutator is
\begin{widetext}
\begin{equation}
  \label{eq:comm-bdag}
  \bigl[\mathcal{H}^{(2,0)}(s),b_j^{\dagger}\bigr]
  =
  \begin{cases}
    (1-s)\bigl(a_0^{\dagger}+\delta\,a_0\bigr), & j=0,\\[4pt]
    (1-s)\bigl(a_j^{\dagger}+\delta\,a_j\bigr)
      + s\bigl(a_{j+1}^{\dagger}+\delta\,a_{j+1}\bigr), & j\le -1.
  \end{cases}
\end{equation}
\end{widetext}
Seeking now a ZM described by a linear combination of the form
$B_R=\mathcal{N}^{-1}\sum_{j\le 0}c_j\,b_j^{\dagger}$, one arrives at
\begin{equation}
  \label{eq:BR}
  B_R = \frac{1}{\mathcal{N}}\sum_{j=-\infty}^{0}
        b_{j}^{\dagger}\left(-\frac{1-s}{s}\right)^{\!-j},
\end{equation}
which commutes with $\mathcal{H}^{(2,0)}$, as does $B_R^{\dagger}$. With the
same normalization $\mathcal{N}^{2}=s^{2}/(2s-1)$, the
commutator of these two ZMs reads
\begin{equation}
  \label{eq:normBR}
  \bigl[B_R^{\dagger},B_R\bigr]
  = \frac{1}{\mathcal{N}^{2}}\sum_{n=0}^{\infty}
    \left(\frac{1-s}{s}\right)^{\!2n}
  =
  \begin{cases}
    1 & s>1/2,\\
    \infty & s<1/2,
  \end{cases}
\end{equation}
so that, exactly as for the left-boundary half-chain, the right-boundary ZMs
form a conjugate pair of canonical bosons only in the $s>1/2$ region of the
dynamically stable phase.
}

The analytic form of the ZMs in Eqs.\,\eqref{al_mode} and \eqref{eq:BR} immediately
yields a closed-form expression for the localization length. By
rewriting the expansion coefficients as an exponential, one finds that
\begin{align}
\label{eq:LocLengthAnalytic}
    \xi(s)=\frac{1}{\log\!\left(\frac{s}{1-s}\right)}.
\end{align}
This expression analytically confirms the two limiting behaviors observed numerically in Fig.~\ref{fig:Localization}: $\xi\rightarrow\inf$ as $s\rightarrow 1/2^{+}$ and $\xi\rightarrow 0$ as $s\rightarrow 1^{-}$ (the ZMs are perfectly localized at $s=1$). {The localization length of the 
ZMs analytically determined from the semi-infinite chains fits the numerically extrapolated one from the corresponding OBC ZMs extremely well.} Remarkably, the localization length in Eq.~\eqref{eq:LocLengthAnalytic} depends only on the interpolation parameter $s$ and is entirely independent of the pairing parameter $\delta$, the very parameter that drives the dynamical instability. This is a strong hint as to the independence of the topological phase transition at $s=1/2$ from the dynamical stability one controlled by $\delta$: the spatial structure of the edge modes is a property of the band topology alone, wholly insensitive to how close the system is to the onset of dynamical instability.

\section{Correlation functions and bulk criticality}

{In topological phase transitions, as the ones found in systems of non-interacting fermions, band-gap closings at the Fermi energy are accompanied by criticality -- namely, the development of long-range correlations in the many-body ground state~\cite{Hastings}.} One might therefore expect a similar phenomenon for free bosons. The situation is complicated, however, by the fact that the interpolation model is \textit{thermodynamically} unstable throughout the parameter regime we consider: the Hamiltonian is not bounded from below, so no ground state exists. Nevertheless, as shown in Ref.\,\onlinecite{Criticality}, the QPV (i.e., the state annihilated by all Bogoliubov quasiparticle operators), serves as a natural substitute for the ground state whenever the QBH is dynamically stable, for the purpose of investigating correlation functions. It is not unusual for the QPV to be unique in the bulk; a sufficient spectral condition was provided in Ref.~\onlinecite{Criticality}
in terms of the \rev{Krein gap defined in Eq.~\eqref{KG}.}
If the QPV is not unique, one needs to investigate the behavior of correlations on the whole manifold of the QPV states. 

Specializing this discussion to our interpolation model, as we observed at the end of Sec.~\ref{sec:dynamical-phases}, the Krein gap is identically zero for all values of the parameters for which dynamical stability holds. The family of eigenvectors of Eq.\,\eqref{eq:family} parametrizes a family of distinct QPVs. \rev{In the language of continuous-variable quantum information~\cite{CVQI}, a  QPV is a zero-mean pure Gaussian state, fully characterized by its {\it covariance matrix }(CM):}
\begin{align}
(\mb{C}_{j,j+r})_{\ell\ell'} \equiv \braket{\{\phi_{j,\ell} - \langle \phi_{j,\ell}}, \phi_{j+r,\ell'}^\dag - \braket{\phi_{j+r,\ell'}^\dag} \} \rangle,
\end{align}
where $\phi_j\equiv [a_j,b_j,a_j^{\dag},b_j^{\dag}]^T$. With this notation in place,
the CM of the family of QPVs obtained from Eqs.\,\eqref{eq:family}-\eqref{eq:family2} 
can be calculated, in momentum space, by way of the formula\cite{Criticality}
\begin{align}
    \mb{C}(k;\eta,\theta) &= \mb L(k;\eta,\theta)\,\mb L(k;\eta,\theta)^{\dag}, \\
    \mb L(k;\eta,\theta) &= \begin{pmatrix}
        \vec{u}_{1,+} & \vec{u}_{2,+} & \vec{u}_{1,-} & \vec{u}_{2,-}
    \end{pmatrix}. \nonumber
\end{align}
The Fourier transform of $\mb C(k)$ yields two-point correlations in real space. 
After substituting Eqs.\,\eqref{eq:family}-\eqref{eq:family2} into the general expression of the CM, dramatic 
simplifications occur, and the CM boils down to the following closed-form expression:
\begin{widetext}
\begin{align}
\label{eq:Gamma-eta}
&\mb{C}(k;\eta,\theta) = \\
&\left(
\begin{array}{cccc}
 \cosh (2 \eta(k) ) & 0 & -e^{-i \theta(k) } \sinh (2 \eta(k) ) & 0 \\
 0 & \frac{\left(\delta ^2+1\right) \cosh (2 \eta(k) )-2 \delta  \sinh (2 \eta(k) ) \cos (\theta (k))}{1-\delta ^2} & 0 & \frac{-2 \delta 
   \cosh (2 \eta(k) )+\left(\delta ^2 e^{i \theta(k) }+e^{-i \theta (k)}\right) \sinh (2 \eta(k) )}{1-\delta ^2} \\
 -e^{i \theta(k) } \sinh (2 \eta(k) ) & 0 & \cosh (2 \eta (k)) & 0 \\
 0 & \frac{-2 \delta  \cosh (2 \eta (k))+\left(\delta ^2 e^{-i \theta(k) }+e^{i \theta (k)}\right) \sinh (2 \eta )}{1-\delta ^2} & 0 &
   \frac{\left(\delta ^2+1\right) \cosh (2 \eta (k))-2 \delta  \sinh (2 \eta(k) ) \cos (\theta(k) )}{1-\delta ^2} \\
\end{array}
\right). \notag
\end{align}
\end{widetext}

Several features of Eq.~\eqref{eq:Gamma-eta} are noteworthy. 
First and foremost, the CM is completely {\it independent} of the
interpolation parameter $s$. Since $s$ is the parameter that drives
the topological phase transition, this fact strongly suggests that, whatever they may be, the \textit{QPV correlations are not in any way linked to this transition}. In other words, for the interpolation model, the correlations in any one $s$-dependent family of QPVs are blind to the band topology and boundary physics. 
Second, the $k$-dependence of the CM is inherited entirely from the free parameters $\eta(k), \theta(k)$, which are unconstrained precisely because the Krein gap is identically zero. On the one hand, we can thus choose these functions so as to elicit long-range correlations in the selected QPV. This is consistent with the general analysis of Ref.~\onlinecite{Criticality}, where the vanishing of Krein gap 
was identified as the necessary condition under which long-range QPV
correlations are permitted to occur. On the other hand, the QPV correlations 
are exponentially bounded if and only if the CM entries are analytic in the 
Brillouin zone. Accordingly, for $\eta(k), e^{i\theta(k)}$ periodic and analytic 
in the Brillouin zone, correlations are exponentially bounded. 

\rev{Physically, each choice of $(\eta(k),\theta(k))$ corresponds to a different
pure, zero-mean Gaussian state. As such, it can be accessed in two ways. The first is through unitary control resources, by
two-mode squeezing the Fock vacuum, as available in cavity- and circuit-QED
and optomechanical arrays~\cite{2024Slim,2024Busnaina}. The second is
dissipative, with the target state arising as the unique steady state of a suitably engineered  
Markovian Gaussian master equation described by a quadratic bosonic
Lindbladian (QBL)~\cite{GaussianSS,Bosoranas,PostBosoranas}. 
Since every QPV is a degenerate eigenstate of the QBH of interest, in
practice the choice is fixed by the preparation. Different choices then lead
to genuinely distinct states, with distinct CMs and observable QPV
correlations.}

What is the take-away message? Long-range correlations are available in the interpolation model, but their presence or absence is governed entirely by the choice of QPV, a freedom granted by the uniformly vanishing Krein gap, not by the value of $s$. The possibility of preparing the system in a critical QPV persists even at $s=0$ and $s=1$, where the chain is a set of decoupled dimers. This almost accidental availability of critical QPVs at the topological transition line is in contrast with the model's sharply structured and predictable topological bulk-boundary behavior. To this extent, bosonic criticality and topological transitions are decoupled features of the interpolation model. 

In addition to the topological transition as a function of $s$, the interpolation model also supports a dynamically unstable regime for $\delta>1$. Generically, a transition from \rev{dynamical} stability-to-instability is the precursor of long-range correlations in the QPV of a QBH \cite{Criticality}. For the interpolation model, the non-zero elements of the CM diverge at $\delta=1$ regardless of $(\eta, \theta)$. The divergence of the CM sits at the dynamical stability phase boundary where the dynamical matrix develops EPs, which, as established in Ref.~\onlinecite{Criticality}, is generically associated with signatures of criticality. With this critical transition at $\delta=1$, it would be useful to know if there is also an associated change in the topological structure of the interpolation model. We will argue next that the answer is in the negative: the topological bulk-boundary physics is indeed insensitive to the value of $\delta$ as the boundary ZMs suggests. This provides further indication that, at least in our model, QPV criticality and topological transitions are independent features since there is no known sensible notion of a QPV for a dynamically unstable QBH.

\section{Topological physics beyond dynamical stability}
\label{sec: beyondstable}

\rev{The boundary-localized ZMs, identified in Sec.\,\ref{sec:boundary} and present for $s>1/2$, persist independently of the value of $\delta$. Thus, it would be valuable to know, in the most unambiguous terms possible, whether the transition at $s=1/2$ is truly topological, both in the dynamically stable and unstable phases.} Thus, topological tools that bypass the spectral consideration of dynamical stability would be ideal companions at this point in the investigation. \rev{(For an alternative approach, restricted to the dynamically stable regime, see Appendix~\ref{sec: BerryPhase})}.
 
Taking inspiration from the SSH model, we zoom in on the chiral pseudo-symmetry $\mb R$ of the interpolation model and investigate the full class of bosonic dynamical matrices that anti-commute with $\mb R$. Interestingly, even though we are not requiring particle-number conservation, we will find out that this bosonic chiral pseudo-symmetry class is remarkably similar in structure, if 
not in the details of how the structure emerges, to the particle-number conserving fermionic BDI class as described in terms of many-body 
symmetries\cite{2016Kennedy, Alldridge2020, 2023Alase}.
(Our analysis here is inspired by our previous work with fermions of Ref.~\onlinecite{2023Alase}.) The bosonic class determined by the chiral pseudo-symmetry $\mb R$ supports a topological classification and a bulk-boundary correspondence in 1D. The bosonic SSH fits well in this class as a representative of a topologically non-trivial subclass, albeit with a differently normalized topological label due to the class being particle-number non-conserving. These remarks hint at the much more detailed investigation of bosonic many-body symmetry classes in Ref.~\onlinecite{gardin2026manybodysymmetryprotectedzeroboundary}, to which we refer the interested reader for more details. Our focus here is the interplay of topology and bosonic (generalized) criticality.

\subsection{A bosonic chiral pseudo-symmetry class}

Let us pin down the structure of a general bosonic dynamical matrix in the class of the pseudo-symmetry $\mb R$, satisfying $\mb R \mb G \mb R =-\mb G.$ In addition, as befits a chiral pseudo-symmetry, $\mb R$ is a unitary matrix that squares to the identity. Space dimensionality is beside the point at this level of analysis. 

To bring the role of the chiral pseudo-symmetry to the fore, it is useful to reorganize the bosonic Nambu array as \cite{GBT1}
\begin{equation}
\Phi\equiv [a_1,\ldots,a_N,a_1^{\dag},\ldots,a_N^{\dag},
    b_1,\ldots,b_N,b_1^{\dag},\ldots,b_N^{\dag}]^T.
\end{equation}
Then, Eq.\,\eqref{eq:R-def} takes the rearranged form
\begin{equation}
\label{eq:R-def2}
\mb R \;\equiv\; \mathrm{diag}(-1,-1,+1,+1), 
\end{equation}
and a bosonic dynamical matrix anticommutes with it
if and only if the latter is block off-diagonal. Factoring in 
the usual bosonic requirements on a dynamical matrix, we arrive at the bipartite structure
\begin{equation}
\label{eq:gpartite}
{\mb G}\equiv \begin{bmatrix} 0 & \boldsymbol{\tau}_3\mb B^\dagger \boldsymbol{\tau}_3\\
\mb B & 0
\end{bmatrix} 
\end{equation}
as the general form of a bosonic dynamical matrix in the chiral class defined by $\mb R$, \rev{with $\boldsymbol{\tau}_j$ being} redefined in accordance with the Nambu array rearrangement. $\mb B$ must satisfy the constraint
\begin{align}
\label{eq: tau1condition}
\boldsymbol{\tau}_1\mb B^*\boldsymbol{\tau}_1=-\mb B,
\end{align}
but is otherwise arbitrary.
  
There is a more informative characterization of the 
set of admissible matrices $\mb B$. If 
\begin{equation}
\mb X\equiv \begin{pmatrix}
    \mb r_1 & \mb r_2\\
    \mb r_3 & \mb r_4
\end{pmatrix}
\end{equation}
stands for an arbitrary real matrix, and 
\begin{equation}
\mb M\equiv \frac{1}{\sqrt{2}}\begin{pmatrix}
    1 & i\\
    1 & -i
\end{pmatrix}\otimes \mb 1_N,
\end{equation}
then one can check that 
$\mb B = i \mb M \mb X {\mb M}^{-1}$ satisfies
Eq.~\eqref{eq: tau1condition}. Conversely, expanding $\mb B$ as
\begin{align}
   \mb B=i \mb I_2\otimes \mb a_0+i \boldsymbol{\sigma}_1\otimes \mb a_1+
   i \boldsymbol{\sigma}_2\otimes \mb a_2+ \boldsymbol{\sigma}_3\otimes \mb a_3,
\end{align}
one learns that Eq.~\eqref{eq: tau1condition} is satisfied if and only if $\mb a_i$ 
are all real matrices. With this information in hand, one can check directly that 
$\mb X= -i\mb M^{-1}\mb B\mb M$ is a real matrix.  

At this point, we can make a clearer structural similarity between our bosonic chiral class and the fermionic BDI class (the class of the fermionic SSH model) alluded to earlier. After a suitable 
many-body symmetry analysis 
in the case of fermions\cite{2023Alase}, and the pseudo-symmetry analysis above in the present case of bosons, one concludes that both classes are parametrized by 
the real square matrices.

\subsection{Topological classification and bulk-boundary correspondence in 1D}
\label{subsec:BB}

Let us now zoom in on chains with a clean (no quenched disorder) bulk. In 1D,
our chiral class is topologically non-trivial and supports a bulk-boundary 
correspondence \rev{by way of the Krein-Gohberg index theorem for Toeplitz 
operators\cite{lax2014functional} (see Appendix~\ref{app:KGindexth} for
a quick, descriptive introduction).} Moving to $k$ space, a chiral dynamical matrix takes the form
\begin{align}
\nonumber {\mb g}(k)&= \begin{bmatrix} 0 & \boldsymbol{\tau}_3 \mb b(k)^\dagger \boldsymbol{\tau}_3\\
\mb b(k) & 0
\end{bmatrix}, \quad 
-\mb b(-k)=\boldsymbol{\tau}_1\mb b(k)^*\boldsymbol{\tau}_1,
\end{align}
with $\mb b(k)$ corresponding to the block $\mb B$ of the real-space $\mb G$ 
translated to Fourier space in the usual ways. Our tool of choice, an index theorem, requires that we zoom in on dynamical matrices that satisfy a kind of band-gap condition: for the bulk system, zero should be in a band gap. It is convenient to capture this requirement by imposing that $\det \mb g(k) \neq 0$ for all values of $k$. Since $\det \mb g(k) = |\det \mb b(k)|^2$, it follows that $\det \mb b(k) \neq 0$ for all values of $k$. It is this latter condition that must hold for deploying the Krein-Gohberg index theorem.
Hence, in 1D, we can associate with every translationally-invariant, chiral dynamical matrix satisfying the \textit{Fredholm gap condition} \rev{a topological (homotopy) invariant, an \textit{index}}: 
\begin{equation}
\rev{\text{Ind }} \mb g(k) = \text{winding number}(\det \mb b(k), 0).
\label{index}
\end{equation}
In words, $\det \mb b(k)$ is a closed path in the complex plane, that misses the origin due to the gap condition. The number of times it winds around the origin is a topological invariant \rev{ -- the index -- which we use} to classify the gapped, chiral $\mb g(k)$. 

Having topologically classified these 1D  dynamical matrices in this way, it is now straightforward to recognize the associated bulk-boundary correspondence. At this point, we need to switch from BIBCs \rev{to L- (or R-)SIBCs} to allow for ZMs to enter the spectrum of the dynamical matrix. \rev{Let us first consider the left-edge ZMs by looking at 
$$
\mb G^L \equiv \begin{bmatrix} 0 & \boldsymbol{\tau}_3 \mb B^{L \dag} \boldsymbol{\tau}_3\\
\mb B^L& 0
\end{bmatrix},
$$ 
a chiral bosonic dynamical matrix of the semi-infinite chain that has a left-boundary, corresponding to the BCs imposed by $\mb T^L$ in Eq.~\eqref{LR}.} Zooming in on the block $\mb B^{L}$, what can we say about its ZMs, that is, its kernel? The Krein-Gohberg index theorem tells us that
\begin{eqnarray*}
\dim \ker \mb B^L-\dim\ker \mb B^{L\dagger} 
 = \text{winding number}(\det {\mb b(k)}, 0).
\end{eqnarray*}
In short, there is nothing useful one can say about the ZMs of $\mb B^{L}$. However, thanks to its block off-diagonal structure (from the chiral pseudo-symmetry), it is the case that
$$\dim \ker\, \rev{\mb G^L}
= \dim \ker \mb B^L
+ \dim \ker \mb B^{L \dag}.$$
Thus, one concludes that
\begin{equation}
\dim \ker\, \rev{\mb G^L}
\geq |\rev{\text{Ind }}\mb g(k)|.
\label{bbc}
\end{equation}
This is the basic bulk-boundary correspondence for our chiral bosonic class in 1D: the total number of independent boundary ZMs of $\mb G^L$ cannot be smaller than the topological classifying invariant of the clean bulk system. {An exactly analogous calculation can be performed for right-ZMs, by exchanging $\mb G^L \mapsto \mb G^R$, the latter being the dynamical matrix associated to R-SIBC, and to the operator $\mb T^R$ also defined in Eq.~\eqref{LR}.}

Let us conclude \rev{this section} with a remark on robustness. Without any additional work, we can learn from the index theorem that changing the BCs of a topologically non-trivial chain -- for example, by adding boundary disorder -- may change the number of boundary ZMs but cannot make the number drop below the bound set by the topological invariant, provided \textit{that the modified BC does not break the chiral pseudo-symmetry} and take the system out of its class. Mathematically, changing the BCs amounts to adding a compact perturbation to the dynamical matrix, which should also anti-commute with $\mb R$ for the bulk-boundary correspondence to be unaffected. In this sense the pseudo-symmetry protects the boundary ZMs. It is substantially harder to strengthen this result to incorporate quenched bulk disorder as has been done for fermions \cite{Alldridge2020}, and it is not clear to us how bosonic physics might modify the approaches that work for fermions when there are protecting symmetries. There is a worked-out case in the literature for thermodynamically stable, two-dimensional systems without protecting symmetries; see Ref.~\onlinecite{10.1063/1.5002094}.

\subsection{The boundary between stable regimes is also a topological boundary}

Let us re-examine the interpolation model of Eq.\,\eqref{Eq: InterpolationHamiltonian} in this new light.  The lower-left block of the \rev{relevant dynamical matrix $\mb g^{(2,0)}(k)$} is
\begin{equation*}
    \rev{\mb b^{(2,0)}(k)} =((1-s)+s e^{ik})\begin{bmatrix} 1
    & \delta \\ -\delta & -1 \end{bmatrix}.
\end{equation*}
Since
\begin{equation*}
    \det \mb b^{(2,0)}(k)= ((1-s)+s e^{ik})^2 (-1 +\delta^2),
\end{equation*}
the bands of the dynamical matrix do not touch zero, regardless of whether or not they are real valued, for 
$0\leq \delta<1$ ($\delta =0$ is the bosonic SSH model) 
and $1<\delta<\infty$. Closer inspection reveals that, for 
$\delta\neq 1$ fixed, the bands only touch zero for $s=1/2$. 
By visualizing the path $\det \mb b^{(2,0)}(k)$ in the complex plane, one concludes that 
\begin{align}
   \rev{\text{Ind}}\ \mb g^{(2,0)}(k)
    =\begin{cases}
    2 & s>1/2 \ \text{and}\  \delta\neq 1 ,\\
    0 & s<1/2\ \text{and} \ \delta\neq 1,
\end{cases}
\end{align}
\rev{implying there are at least 2 ZMs on the left edge, independently of $\delta<1$ or $\delta>1$, for $s>1/2$. Thus, we confirm that both the dynamically stable and unstable regions split into two phases -- topologically trivial and non-trivial. Complemented by an equivalent computation under R-SIBCs, the bulk-boundary correspondence thus predicts at least two ZMs per edge in the topologically non-trivial phase. This is consistent with our investigation of the boundary modes of the interpolation model in Sec.~\ref{sec:boundary}, where, for $s>1/2$, independently of $\delta$, we find two canonically conjugate ZMs on each boundary.} We also know that the ZMs happen to persist at the boundary of dynamical stability,
but those are not isolated from the bulk modes 
or mandated by the index theorem for Toeplitz operators.

\section{Discussion and outlook}
\label{sec:discussion}

For systems of free bosons, long-range correlations are expected to arise in regions of parameter space
that separate a dynamically stable regime from a dynamically
unstable one~\cite{Criticality}. This creates a conceptual challenge: namely, to identify qualitatively different phases inside a dynamically stable regime -- which is often considered more physically relevant -- even if one cannot, as a rule, expect these phases to be separated by a critical boundary. Topological band structure is a natural candidate for \rev{tackling this} challenge. However, this idea immediately raises a puzzling objection: \rev{on the one hand,} for fermions, boundaries separating distinct topological phases must be critical. \rev{On the other hand,} for dynamically stable bosons, criticality is, at best, a fine-tuned occurrence -- it can only occur at the boundaries of dynamical stability. In an effort to understand the situation better, in
this paper we introduced a quadratic bosonic chain, an ``interpolation model,'' that illustrates in concrete terms the scenario suggested above: in the dynamically stable regime, \rev{we uncover} two topologically distinct phases. We base this conclusion on two different approaches \rev{which, in this regime, yield consistent results. In the dynamically unstable regime, only one of the two approaches remains viable.}

The line separating the two topological phases and fully contained within the stable regime can be characterized by the closing of the band gap. From the point of view of its bulk correlations, the behavior of the interpolation model on this line is most peculiar: It hosts a degenerate manifold of QPVs, and the correlations in this manifold can be short- or long-range depending on the choice of QPV. Thus, one may describe the topological transition line in the dynamically stable regime as ``conditionally critical''. In fact, due to the Krein gap being zero throughout the dynamically stable region, at every point in this region there is a family of conditionally critical QPVs, which are independent of the parameter that drives the topological transition. Hence, criticality and topology decouple fully. By contrast, the bulk-boundary behavior of our model satisfies all the usual expectations one may carry over from topologically non-trivial free-fermion phases. 

In addition \rev{to the above transition line}, there is a line separating the dynamically stable regime, and the topological phases in it, from the dynamically unstable regime. This line is, as expected~\cite{Criticality}, critical in the QPV, as it is composed of EPs. \rev{Further to that,} it is also a boundary line for the topological phases -- \rev{the Fredholm gap condition fails along the entire $\delta=1$ line. Consequently, although the topological index happens to take the same value on either side of $\delta=1$, the dynamically stable and unstable regions at fixed $s$ are distinct topological phases, since no gap-preserving path connects them.} We confirm that the topological phases and transition reappear in the dynamically unstable regime, where there is no QPV or known suitable substitute for investigating correlations. 

Putting it all together, we reach the following preliminary conclusions. First, it is indeed possible to differentiate topological phases {\it both} inside the dynamically stable and dynamically unstable regimes. Second, for the topological phases inside the dynamically stable regime, the correlations at the topological phase transition can be \textit{short-range} or {\it conditionally long-range}. The take-home message is that, for \rev{free} bosons, one may well observe a topological phase transition by way of the drastic re-arrangement of the boundary physics, with no criticality in the bulk to go with it. No such behavior is possible for fermions.   

Let us also highlight a couple of interesting open questions. Is it possible for a bosonic topological phase to include both dynamically stable and unstable regions?  If this is indeed possible (and we know of neither an obstruction nor an example), then the topological phase would also contain one or more critical boundaries in it.  One would be able to observe transitions in and out of bulk criticality with no observable impact on the boundary physics. 
It is a converse scenario to the conditionally-critical topological transition we document in this paper. Likewise, is there some sensible way to investigate correlations at a topological phase transition in the dynamically unstable regime? The obstacle here is that the structure of a dynamically unstable QBH fails to single out any many-body state as ``special" in a manner similar to how the QPV stands out for a dynamically stable QBH. 

\rev{In closing, we call attention to a remarkable parallel between topology in QBHs and QBLs which, as mentioned, describe Markovian non-interacting bosonic dynamics. In Refs.\,\onlinecite{Bosoranas,PostBosoranas}, it was demonstrated that the band topology of the dynamical matrix of a QBL need not be reflected in the structure of its {\it bulk} steady-state, which in fact need not exist altogether. While signatures of non-trivial {\it non-Hermitian} topology may still be encoded in both two-time \cite{PostBosoranas} and equal-time \cite{PorrasLongRange} correlations under OBCs, the appearance of topologically mandated (Majorana or Dirac) bosonic ZMs is only supported in a {\it transient} metastable regime. Likewise, in our model QBH, we find that the topological transition and the appearance of boundary ZMs is entirely divorced from the structure of the QPV and the emergence of long-range correlations {\it in the bulk}.} What deeper significance such a parallel may carry remains to be better understood. We hope to return to these questions in the future.

\section*{Acknowledgment}
It is a pleasure to thank Vincent P. Flynn for many insightful discussions of bosonic physics over the years and for a critical reading of the manuscript. L.V. gratefully acknowledges support from the US National Science Foundation through Grant No.\,PHY-2412555.

\appendix

\section{Band topology in the dynamically stable regime}
\label{sec: BerryPhase}

In the dynamically stable regime, the usual Berry's phase of Bloch eigenfunctions, as extensively used for topological free-fermion physics~\cite{Kane,Deng}, can be modified to account for the symplectic structure of the bosonic dynamical matrix to continue to yield quantized numbers~\cite{Peano2015}. The appropriate inner product for the symplectic variant of the Berry phase is the indefinite inner product $\vec{v}^{\dag}\boldsymbol{\tau}_3\vec{v}$ from before. Our setting differs from the canonical one in one crucial aspect, however. Since the Krein gap vanishes throughout the dynamically stable regime, every band of the dynamically stable matrix is two-fold degenerate. Therefore, the symplectic Berry phase is not a function of the band alone, but of the choice of free parameters in Eq.\,\eqref{eq:family}. Evaluated on the base eigenvectors $\vec{v}_{j,\pm}$, it is quantized and jumps in value from $s<1/2$ to $s>1/2$, matching the boundary physics of Sec.\,\ref{sec:boundary}. However, for arbitrary choices of $\theta(k),\eta(k)$, the quantization is lost. For this reason and the fact that the calculation of the Berry phase breaks down in the dynamically unstable regime, the index-theoretic characterization of the topological phases we present in Sec.\,\ref{sec: beyondstable} is the more robust of the two. We nonetheless include this supplementary Berry-phase picture here as it provides a more physically transparent bulk diagnostic, directly parallel to the fermionic story and, for $\vec{v}_{j,\pm}$, it reproduces the same topological characterization as the index theorem throughout the dynamically stable regime.

The symplectic Berry phase \rev{for the $j$th band of the interpolation model} is given by:
\begin{align}
\label{eq: berryphase}  
  \gamma^{\pm}_j=\pm i\int^{\pi}_{-\pi}dk
        \left(\vec{v}_{j,\pm}^{\dag}(k) \,\boldsymbol{\tau}_3 
        \frac{d \vec{v}_{j,\pm}(k)}{dk} \right).
\end{align}
 Letting $k_{\ell+1}=-\pi+\frac{2\pi \ell}{N}$, $\gamma^{+}_j$ can be
computed in a numerically gauge-invariant way as~\cite{Squaring}
\begin{align}
\label{eq:GaugeInvariant}
   \gamma^{+}_j=\underset{N\rightarrow\inf}{\lim}\text{Im}\log
   \prod\limits_{\ell=1}^{N}
   \left(\vec{v}_{j,+}^{\dag}(k_{\ell+1}) \boldsymbol{\tau}_3
         \vec{v}_{j,+}(k_{\ell})\right).
\end{align}
Evaluating Eq.~\eqref{eq:GaugeInvariant} as a function of $s$ yields a sharp dichotomy. For $0 \leq s < 1/2$, the symplectic Berry phase vanishes for every band, $\gamma_j^{\pm} = 0$, indicating that the interpolation model is topologically trivial in this regime. For $1/2 < s \leq 1$, on the other hand, one finds $\gamma_j^{\pm} = \pm\pi$, so that the corresponding symplectic invariant takes the nontrivial value $\vert \gamma_j^\pm / \pi\vert_{\text{mod}(2\pi)} = 1$. This suggests that the transition of the interpolation model at $s=1/2$ (and for $\delta<1$) is topological, that is, associated with a jump of a band topological invariant. Both this method and the index theoretic method discussed in Sec. \ref{subsec:BB} identify identical regions as topologically trivial or non-trivial, within the dynamically stable regime.

\section{The index theorem of Krein and Gohberg in a nutshell}
\label{app:KGindexth}

This appendix is meant to be a friendly, descriptive 
introduction to the Krein-Gohberg index theorem that informs our bulk-boundary correspondence of Sec.~\ref{subsec:BB}. Our presentation is adapted from the textbook Ref.~\onlinecite{lax2014functional}, but we take a different sign convention for consistency with the physics literature. For a more advanced treatment, see Ref.~\onlinecite{bleecker2013index}.

A translationally-invariant operator on the bi-infinite lattice is determined by its \textit{symbol}, a continuous, $2\pi$-periodic, matrix-valued function
\begin{equation}
\mathbf{b}(k)=\sum_{n\in\mathbb{Z}}e^{ink}\mathbf{b}_n,
\qquad
\mathbf{B}=\sum_{n\in\mathbb{Z}}\mathbf{V}^{\dagger n}\otimes \mathbf{b}_n ,
\end{equation}
the unit-cell matrices $\mathbf{b}_n$ being recovered from $\mathbf{b}(k)$ by Fourier inversion.

The next step is to map the translationally invariant operator $\mathbf{B}$ associated to $\mb b(k)$ to the corresponding operator for a half-infinite lattice with one termination on the left. This is done simply by replacing the shift operators appropriate for BIBCs with those appropriate for L-SIBCs, recall Eqs.~\eqref{Sshift}-\eqref{LR} (an analogous treatment can be done for the half-chain with a right termination, by instead considering R-SIBCs). The resulting operator,
$$
\mb B^L=\sum_{n\in\mathds{Z}} \mathbf{T}^{L\dagger\, n} \otimes \mb b_n, $$
is known in the mathematical literature as a {\it block-Toeplitz operator}; it is block-constant along diagonals. 

$\mb B$ is invertible when $\det\mathbf{b}(k)\neq0$ for all $k$. However, this does not survive once boundaries are imposed. $\mb T^L$ annihilates the boundary site. What does survive is that $\mathbf{B}^L$ is \textit{Fredholm}, meaning
$\ker\mathbf{B}^L$ and $\ker\mathbf{B}^{L\dagger}$ (the latter being the
cokernel of $\mb B^L$) are both finite-dimensional, so that the integer
\begin{equation}
\mathrm{index} \,(\mathbf{B}^L)\equiv\dim\ker\mathbf{B}^L-\dim\ker\mathbf{B}^{L\dagger}
\end{equation}
is well-defined. More precisely, it is 
invariant under homotopy, meaning that, if $\mb B^L_t$ is a continuous path of Fredholm operators, then the index is the same integer for every value of $t$, that is, everywhere along the path. One concludes that the topological space of Fredholm operators breaks up into a countable infinity of path-connected components, and the components are labeled by the index.

Now, as we mentioned, we can decide whether a Toeplitz operator is also a Fredholm operator by studying the symbol: the Fredholm condition can be recast as $\det\,\mb b(k)\neq 0$. Since, in addition, $\det\,\mb b(k)$ traverses a closed loop in the complex plane as $k$ grows from $k=-\pi$ to $k=\pi$, there is another integer one can calculate:
the winding number of $\det\,\mb b(k)$ around the origin of the complex plane. This integer is also a homotopy invariant. The remarkable theorem discovered by Krein and Gohberg~\cite{lax2014functional}, 
with hindsight, the simplest special case of the Atiyah-Singer index theorem~\cite{bleecker2013index}, is that {\it these two integers agree}. As a formula, 
\begin{equation}
    \label{ToeplitzITh}
    \mbox{Ind}\, (\mb B^L) = 
    \mbox{winding number}(\det \mb b(k), 0). 
\end{equation}

While our discussion so far puts all the emphasis on OBCs, even basic index theorems like the Gohberg-Krein example accommodate arbitrary BCs from the outset, in the following sense. Changing the BCs or adding an impurity at the edge modifies only a finite number of matrix elements, all of them near the boundary. Such a modification is a finite block matrix, and any finite matrix can be viewed as an operator on the half-infinite lattice simply by letting it act as zero on
all the remaining sites. The defining characteristic of this set of operators is that their range is finite-dimensional. The closure of the set of operators of finite range, in the operator-norm topology, is the algebra of compact operators. 

The range of a compact operator is no longer necessarily finite-dimensional. However, it is still true that a compact operator decays into the bulk, so that its action becomes indistinguishable from that of the zero operator far enough from the boundary, up to some tolerance. The smaller the tolerance, the further away from the boundary one must move. In this sense, if $\mb C$ is a compact operator, then one can think of $\mb B^{L'}=\mb B^L+\mb C$ as the same operator up to a change of BCs or addition of an impurity on the boundary. This observation is useful because the analytical index is unaffected by such a modification, that is, 
$\mbox{index}\, \mb B^L= \mbox{index}\, \mb B^{L'}$. 
This kind of robustness against boundary perturbations can also be tested numerically with proper attention to finite-size scaling.

\bigskip


%

\end{document}